\documentclass[twocolumn,groupedaddress,amsmath,amssymb,prb,aps,longbibliography]{revtex4-1}

\usepackage{float}
\usepackage{siunitx}
\usepackage{graphicx}
\usepackage{dcolumn}
\usepackage{bm}
\usepackage{booktabs}
\usepackage{multirow}
\newcommand{\dbs}[1]{\underline{\underline{\mathbf{#1}}}}
\usepackage{rotating}
\usepackage{xcolor}

\begin{document}

\title{Machine learning Landau free energy potentials}

\author{Mauro Pulzone,$^{1,2}$ Natalya S. Fedorova,$^{1}$ Hugo
  Aramberri,$^{1}$ and Jorge \'I\~niguez-Gonz\'alez$^{1,2}$}

\affiliation{
  \mbox{$^{1}$Luxembourg Institute of Science and Technology (LIST),
    Avenue des Hauts-Fourneaux 5, L-4362 Esch/Alzette, Luxembourg}\\
 \mbox{$^{2}$Department of Physics and Materials Science, University
   of Luxembourg, Rue du Brill 41, L-4422 Belvaux, Luxembourg}}

\begin{abstract}
We show how to construct Landau-like free energy potentials using a
machine-learning approach. For concreteness, we focus
on perovskite oxide PbTiO$_{3}$, representative of a large class of
materials that undergo non-reconstructive structural phase
transitions; in this case, a proper ferroelectric transformation with
improper ferroelastic features. We work with a training set obtained
from Monte Carlo simulations based on an atomistic
``second-principles'' potential for PbTiO$_{3}$. We rely exclusively
on data that would be experimentally accessible -- i.e.,
temperature-dependent polarization and strain, both with and without
external electric fields and stresses applied --, to explore scenarios
where the training set could be obtained from laboratory
measurements. We introduce a scheme
that allows us to identify optimal polynomial models of the
temperature-dependent free energy surface, mapped as a function of the
homogeneous electric polarization and homogeneous strain. Typically,
our method evaluates thousands of possible models, ranking them by
accuracy and predictive power. Our results for PbTiO$_{3}$ show that a
very simple polynomial -- where only two parameters depend linearly on
temperature -- is sufficient to yield a correct description of the
material's behavior. We thus validate the usual approximations made in
phenomenological studies of phase transitions of this
kind. Remarkably, the obtained models also capture the subtle
couplings by which elastic strain controls key features of
ferroelectricity in PbTiO$_{3}$ -- i.e., the symmetry of the polar
phase and the discontinuous character of the transition --, despite
the fact that no effort was made to include such information in the
training set. We emphasize the distinctive aspects of our methodology
(which relies on an original form of validation step) by comparing it
with the usual machine-learning approach for model construction. Our
results illustrate how physically motivated models can have remarkable
predictive power, even if they are derived from a limited amount of
data. We argue that such ``third-principles'' models can be the basis
for predictive macroscopic or mesoscopic simulations of ferroelectrics
and other materials undergoing non-reconstructive structural
transitions.
\end{abstract}

\maketitle


\section{Introduction}
Machine-learned interatomic potentials (MLIPs) or force fields (MLFFs)
derived from Density Functional Theory (DFT) calculations are
transforming the field of computational materials science. Various
MLIP families already enable an accurate and efficient description of
interatomic interactions, which in turn makes it possible to run
nanoscale molecular dynamics simulations ($\sim$nm, $\sim$ns) in a
customary manner. While many challenges remain ahead -- e.g., to
tackle problems that require an explicit treatment of electrons or
magnetism --, progress in the past decade has been outstanding, and
today several MLIP schemes compete to become the standard approach for
performing most DFT
studies.\cite{bartok15,wang18,jinnouchi19,fan21,musaelian23,xie23,batatia24,jacobs25}

However, whenever we are dealing with mesoscale or macroscopic
phenomena, we are typically interested in length and time scales of at
least $\mu$m and $\mu$s, respectively, where simulations based on
atomistic MLIPs are unfeasible because of their high computational
cost. The theoretical approach to such problems typically relies on
continuum field theories, and even on macroscopic models whenever the
inhomogeneities in the system do not require an explicit
consideration. Of particular interest are cases involving ferroic
materials (ferromagnetic, ferroelectric, ferroelastic, and all related
systems) that often present complex multidomain states that can
nevertheless be treated -- in a qualitative and quasi-quantitative
satisfactory way -- by relatively simple Ginzburg-Landau (continuum)
or Landau (macroscopic)
theories.\cite{hu98,chen07,toledano-book1987,chandra07}

Ginzburg-Landau potentials for ferroics have traditionally been
obtained by fitting to experimental data. In the case of
ferroelectrics, which is our focus here, the homogeneous part of the
potential (i.e., the Landau model) can be derived from the temperature
dependence of the polarization and dielectric
permittivity.\cite{devonshire54,haun87} Then, constructing a Ginzburg-Landau model
requires additional information about the energy cost of having
spatial inhomogeneities in the electric polarization, which can be
deduced, e.g., from the spatial extension of domain
walls.\cite{tagantsev-book2010,meier-book2020} Critically, such potentials capture
thermal effects -- i.e., the temperature evolution of the relevant
free energy landscape -- in a way that is approximate but sufficient
for many purposes.

There have been efforts to derive effective free energy potentials
from first principles, as needed to run predictive mesoscopic
simulations whenever experimental information is not available (e.g.,
for hypothetical materials). For example, several authors have
proposed schemes to fit temperature-dependent Landau potentials using
data from statistical calculations based on atomistic models derived
from DFT.\cite{iniguez01,kumar10} Attempts have also been made to fit
simple temperature-dependent Ginzburg-Landau potentials.\cite{kumar10}
More complex Ginzburg-Landau potentials have also been derived from
DFT, but in this case the existing applications are restricted to the
limit of 0~K as far as we know.\cite{schiaffino17,shapovalov23} While
valuable, these efforts are somewhat limited. Indeed, we still lack a
general and automatic scheme to create free energy potentials that
capture, with an accuracy that can be systematically improved, the
temperature-dependent behavior of materials with arbitrarily complex
interactions.

To do this, one could adopt a generic machine-learning (ML)
approach. For example, if we wanted to learn the behavior of the
macroscopic polarization and macroscopic strain in a ferroelectric, we
could train a neural network that takes as input the environmental
conditions (temperature, external pressure, external electric field)
and produces as output the properties of interest (equilibrium
polarization and strain). Such a scheme might not allow us to
understand the underlying physical mechanisms responsible for the
observed behavior, but it would enable accurate calculations of the
properties of interest within the limits -- of temperature and
external fields -- defined by the training set (TS).

However, in this context, it seems unwise to ignore the mentioned
time-tested application of Landau-like free energy models, which has
enabled a successful treatment of many intricate cases. With this in
mind, here we explore a physics-informed ML approach that builds on
such a key insight, i.e., that polynomial models of the free energy
usually work very well to describe ferroelectricity and related or
analogous phenomena. Our scheme automatically determines effective
potentials capable of reproducing a TS of relevant thermodynamic
information, deriving directly from the data which monomials (that is,
coupling terms) need to appear in the Landau model. As we will show,
not only does the proposed scheme reproduce accurately the data in the
TS, but it also reveals covert couplings that yield non-trivial
behaviors of the compound we investigate here as a relevant example,
ferroelectric perovskite PbTiO$_{3}$.

The manuscript is organized as follow. In Section~II we introduce the
main ideas and formalism of the proposed machine-learning approach. In
Section~III we briefly justify the choice of PbTiO$_{3}$ as a relevant
model material, introducing some of the non-trivial behaviors it
presents. Here we also describe the second-principles calculations
from which we obtain the training set. In Section~IV we discuss the
results of our fitting procedure; we first consider simple models that
depend on the electric polarization alone, and then move to models
that depend explicitly on both polarization and strain. In Section~V
we emphasize the non-trivial physical effects captured by our models,
which highlight their predictive power. Here we also discuss how the
present method relates to more traditional approaches in the ML
model-building literature, emphasizing peculiarities that could be
useful in other contexts. In Section~V we also outline our envisioned
path toward what we term ``predictive third-principles
simulations''. We conclude in Section~VI with a brief summary.

\begin{table}[h!]
\caption{Symmetry-invariant couplings, involving polarization $\bm{P}$
  and strain $\bm{\eta}$, that we consider in
  the Landau free energy potential of PbTiO$_{3}$. These invariants
  are the same for any perovskite compound provided one takes the
  ideal cubic phase ($Pm\bar{3}m$) as reference. Note that the
  $A_{\beta}$ coefficients for the $f_{\beta}(\bm{P})$ terms could
  correspond to a full model $F({\bm P},\bm{\eta};\bm{\mathcal
    E},\bm{\sigma},T)$ or to a renormalized model $\tilde{F}({\bm
    P};\bm{\mathcal E},T)$; in the latter case, they would be written
  with a tilde, as $\tilde{A}_{\beta}$ (see text).}
\vskip 2 mm
\begin{tabular}{ll}
  \hline\hline
  \addlinespace[2ex]
  $A_{\beta}$ & $f_{\beta}(\bm{P})$ \\ \hline 
  \addlinespace[1ex]
       $A_{2i}$     &$P_{x}^{2} + P_{y}^{2} + P_{z}^{2}$ \\
       $A_{4i}$     &$(P_{x}^{2} + P_{y}^{2} + P_{z}^{2})^{2}$ \\
       $A_{22}$     &$P_{x}^{2} P_{y}^{2} +  P_{x}^{2} P_{z}^{2} +  P_{y}^{2} P_{z}^{2}$ \\
       $A_{6i}$     &$(P_{x}^{2} + P_{y}^{2} + P_{z}^{2})^{3}$ \\
       $A_{42}$     &$P_{x}^{4}(P_{y}^{2} + P_{z}^{2}) + P_{y}^{4}(P_{x}^{2} + P_{z}^{2}) + P_{z}^{4}(P_{y}^{2} + P_{x}^{2})$ \\
       $A_{222}$    &$P_{x}^{2} P_{y}^{2} P_{z}^{2}$ \\
       $A_{8i}$     &$(P_{x}^{2} + P_{y}^{2} + P_{z}^{2})^{4}$ \\
       $A_{422}$    &$P_{x}^{4} P_{y}^{2} P_{z}^{2} + P_{x}^{2} P_{y}^{4} P_{z}^{2} + P_{x}^{2} P_{y}^{2} P_{z}^{4}$ \\
       $A_{44}$     &$P_{x}^{4} P_{y}^{4} + P_{x}^{4} P_{z}^{4} + P_{y}^{4} P_{z}^{4}$ \\
       $A_{62}$     &$P_{x}^{6} P_{y}^{2} + P_{x}^{6} P_{z}^{2}  +
  P_{x}^{2} P_{y}^{6} +  P_{x}^{2} P_{z}^{6} + P_{y}^{6} P_{z}^{2} +
  P_{y}^{2} P_{z}^{6}$ \\ 
  \addlinespace[2ex]
      $B_{\beta}$ & $f_{\beta}(\bm{P},\bm{\eta})$ \\ \hline 
  \addlinespace[1ex]
       $B_{12i}$  &$(\eta_{xx} +\eta_{yy} + \eta_{zz})(P_{x}^{2} + P_{y}^{2} + P_{z}^{2})$ \\
        $B_{12}$  &$\eta_{xx} P_{x}^{2} + \eta_{yy}P_{y}^{2} + \eta_{zz}P_{z}^{2}$ \\
        $B_{14i}$  &$(\eta_{xx} +\eta_{yy} + \eta_{zz})(P_{x}^{2} + P_{y}^{2} + P_{z}^{2})^{2}$ \\
        $B_{14}$  &$\eta_{xx}P_{x}^{4}  + \eta_{yy}P_{y}^{4} +  \eta_{zz}P_{z}^{4}$ \\
        $B_{122}$  &$(\eta_{xx} +\eta_{yy} + \eta_{zz}) ( P_{x}^{2} P_{y}^{2} + P_{x}^{2} P_{z}^{2} + P_{y}^{2} P_{z}^{2} ) $ \\
        $B_{22i}$   &$(\eta_{xx} +\eta_{yy} +\eta_{zz})^{2}(P_{x}^{2} + P_{y}^{2} + P_{z}^{2}) $ \\
        $B_{22}$   &$\eta_{xx}^{2} P_{x}^{2} + \eta_{yy}^{2} P_{y}^{2} +\eta_{zz}^{2}P_{z}^{2}  $ \\
        $B_{112i}$   &$(\eta_{xx}\eta_{yy} + \eta_{xx}\eta_{zz} +\eta_{yy}\eta_{zz}) (P_{x}^{2} + P_{y}^{2} + P_{z}^{2}) $ \\
        $B_{1^{'}11}$  &$\eta_{yz}P_{y}P_{z} + \eta_{xz}P_{z}P_{x} +
  \eta_{xy}P_{x}P_{y}$ \\
  \addlinespace[2ex]
   $C_{\beta}$ & $f_{\beta}(\bm{\eta})$ \\ \hline
  \addlinespace[1ex]
        $C_1$      &$\eta_{xx} +\eta_{yy} + \eta_{zz}  $ \\ 
        $C_2$      &$\eta_{xx}^{2} +\eta_{yy}^{2} + \eta_{zz}^{2}$ \\
        $C_{11}$   &$\eta_{xx} \eta_{zz}+ \eta_{yy} \eta_{xx} + \eta_{yy} \eta_{zz}$ \\
        $C_{2^{'}}$ &$\eta_{yz}^{2} + \eta_{xz}^{2}+ \eta_{xy}^{2}$ \\  
  
  \addlinespace[1ex]
  \hline\hline
\end{tabular}
\label{tab:F}
\end{table}
 
\section{Our approach}
\label{sec:approach}

In this section we introduce the Landau treatment of ferroelectric
phase transitions and related effects, taking the case of perovskite
oxides as a representative and relevant example, a generalization
being trivial. We then describe our proposed methodology to learn the
Landau potential from a TS of standard thermodynamic data.

Let us note that here we are interested in describing the free energy
landscape in a wide range of temperatures, and that the vicinity of
the phase transition is not a particular concern for us. For this
reason, and recognizing that Landau potentials were originally
introduced to describe the transition region,\cite{landau-book1980,
  chandra07} we often refer to our free energy models as being
``Landau-like''.

\subsection{Landau-like model of a ferroelectric perovskite}

We start by selecting the three-dimensional polarization vector, ${\bm
  P}$, as the main variable of interest, which we use to describe the
state of the system. We want to study the associated free energy
surface $\tilde{F}=\tilde{F}({\bm P};\bm{\mathcal E},T)$, where $T$ is
the temperature and $\bm{\mathcal E}$ is the external electric
field. The minima of $\tilde{F}$ represent equilibrium states of the
material. In a Landau-like approach, $\tilde{F}$ is written as Taylor
series for the order parameter around a suitable chosen reference
phase with $\bm{P}=\bm{0}$. Formally, we have
\begin{equation}
\tilde{F}(\bm{P};\bm{\mathcal{E}},T) = \tilde{F}_{0}(T) + \sum_{\beta}
\tilde{A}_{\beta}(T) f_{\beta}(\mathbf{P})- \bm{\mathcal{E}} \cdot
\mathbf{P} \;,
\label{eq:F-tilde} 
\end{equation}
where $A_{\beta}(T)$ are temperature-dependent parameters and
$f_{\beta}(\bm{P})$ are polynomial functions that are invariant under
the symmetries of the reference phase. In the case of ferroelectric
perovskites, the reference structure is the ideal cubic phase with
$Pm\bar{3}m$ space group. The corresponding $f_{\beta}$ polynomials
are given in Table~\ref{tab:F} up to the 8th order in the
polarization. Note also the term $\tilde{F}_{0}(T)$, which describes
free energy variations that do not depend on $\bm{P}$ or $\bm{\mathcal
  E}$; we will not discuss it here.

Strain is a key ingredient to describe ferroelectrics, as it
participates in one of their most important properties:
piezoelectricity. Further, many ferroelectrics -- certainly
perovskites like PbTiO$_{3}$ -- are highly sensitive to mechanical
stress.\cite{schlom07} To account for such effects, we need a Landau
potential that includes strains as variables and stresses as
fields. We formally have
\begin{widetext}
\begin{equation}
F (\bm{P}, \bm{\eta} ; \bm{\mathcal E}, \bm{\sigma}, T) = F_{0}(T) +
\sum_{\beta} A_{\beta}(T) f_{\beta}(\bm{P}) + \sum_{\beta}
B_{\beta}(T) f_{\beta}(\bm{P}, \bm{\eta}) + \sum_{\beta} C_{\beta}(T)
f_{\beta}(\bm{\eta}) - \bm{\mathcal{E}} \cdot \mathbf{P} - \bm{\sigma}
\cdot \bm{\eta} \;,
\label{eq:F}
\end{equation}
\end{widetext}
where $\bm{\eta}$ and $\bm{\sigma}$ are the symmetric strain and
stress tensors, respectively; we denote the corresponding tensor
components as $\bm{\eta} = (\eta_{xx}, \eta_{yy}, \eta_{zz},
\eta_{yz}, \eta_{xz}, \eta_{xy})$. Here, $f_{\beta}(\bm{P},
\bm{\eta})$ and $f_{\beta}(\bm{\eta})$ are symmetry-invariant
polynomials analogous to $f_{\beta}(\bm{P})$, while $B_{\beta}(T)$ and
$C_{\beta}(T)$ are their corresponding coefficients. Table~\ref{tab:F}
lists the polarization, strain, and strain-polarization invariants considered in this
work.

It is important to note that the Landau potentials introduced above
are related in the following way:
\begin{equation}
  \tilde{F}(\bm{P};\bm{\mathcal{E}},T) = \min_{\bm{\eta}}
  F(\bm{P},\bm{\eta};\bm{\mathcal{E}},\bm{\sigma}=\bm{0},T) \; .
\end{equation}
This implies that, when we work with $\tilde{F}$, the strain is
treated as a slave variable that follows the polarization. A related
subtlety about Eqs.~(\ref{eq:F-tilde}) and (\ref{eq:F}) is that some
seemingly equivalent coefficients have a tilde in the former but not in
the latter, as they are in fact different quantities. Thus, for
example, we would say that the parameter $\tilde{A}_{\beta}$ is the
{\sl strain-renormalized} version of $A_{\beta}$.

As indicated in the equations above, all the parameters in our models
(e.g., $A_{\beta}(T)$, $B_{\beta}(T)$, etc.) are in principle
$T$-dependent. For a generic coefficient $\phi_{\beta}(T)$, we express
such a dependence as the following Taylor series
\begin{equation}
    \phi_{\beta}(T)= \phi_{\beta}^{(0)} + \phi_{\beta}^{(1)} T +
    \phi_{\beta}^{(2)} T^{2} + \cdots \; ,
\end{equation}
which defines the new $T$-independent coefficients
$\phi_{\beta}^{(n)}$.

Table~\ref{tab:F} lists all the $f_{\beta}$ interaction terms, and
associated $\phi_{\beta}$ parameters, considered in this work. (Some
extensions, to address specific points, are discussed below.) We
truncate the order of the expansion in a way that depends on the
nature of the couplings, allowing for a more detailed description of
the energetics of the primary order parameter, that is, the
polarization $\bm{P}$ in our sample problem. Specifically, for the
terms involving the polarization alone, $f_{\beta}(\bm{P})$, we
consider all symmetry-independent couplings up to the 8th order. For
the strain terms, $f_{\beta(\bm{\eta})}$, we stop at the quadratic
order. Finally, as regards the polarization-strain couplings
$f_{\beta}(\bm{P},\bm{\eta})$, we go up to a relatively high order
($\sim \eta P^{4}$, $\sim \eta^{2} P^{2}$) when the linear strains are
involved, but consider only the lowest-order couplings ($\sim \eta
P^{2}$) for shear strains. Similarly, when it comes to the
$T$-dependence of the coefficients, we consider a more detailed
treatment of those associated to the $f_{\beta}(\bm{P})$ and
$f_{\beta}(\bm{\eta})$ couplings, while we are more restrictive when
treating the $f_{\beta}(\bm{P},\bm{\eta})$ terms. The number of models
considered is thus limited by these truncations and some additional
simplifications -- e.g., we always include the $\tilde{A}^{(0)}_{2i}$
and $\tilde{A}^{(1)}_{2i}$ coefficients in all our models, we always
include the $C^{(0)}_{1}$, $C^{(1)}_{1}$, $C^{(0)}_{2}$,
$C^{(0)}_{2'}$ and $B^{(0)}_{2',11}$ couplings in all our
$F(\bm{P},\bm{\eta})$ models, etc. Furthermore, we truncate the
temperature expansion at the linear term. For the higher-order
couplings ($\sim \eta P^{4}$, $\sim \eta^{2} P^{2}$), we assume no
temperature dependence. Despite these restrictions, we work with a
very large universe of possible potentials: we evaluate over 16~000
inequivalent models of the $\tilde{F}(\bm{P})$ type, while we consider
over 69~000 different $F(\bm{P},\bm{\eta})$ models with up to 17
parameters.

\subsection{Machine learning the Landau potential}

We now discuss our approach to compute Landau free energy
potentials. First, we present how a given model -- as defined by a
specific combination of $\phi_{\beta}^{(n)}$ parameters -- is fitted
to a TS of data. Then, we introduce the way in which we identify
optimal models -- simple, yet accurate and predictive -- from the vast
universe of possible potentials.

\subsubsection{Fitting the parameters of a particular model}

Suppose we are given a training set $\mathcal{D}$ of the form
\begin{equation}
    \mathcal{D} = \left\{ \{ T^{(s)}, \bm{\mathcal{E}}^{(s)},
    \bm{\sigma}^{(s)}; \mathbf{P}^{(s)}, \bm{\eta}^{(s)} \} \right\}\; ,
\end{equation}
where $s = 1, ..., n$ runs over representative equilibrium
configurations. Hence, this TS describes how the state of the system
-- as characterized by its polarization and strain -- depends on the
external conditions of temperature, electric field and stress. For
convenience, in the following we assume that all the quantities in the
TS are scaled in a physically meaningful way, which we describe in
Section~\ref{sec:dimensionless}, so they are dimensionless.

The model $F({\bm P},\bm{\eta};\bm{\mathcal E},\bm{\sigma},T)$ must be
such that the configurations in $\mathcal{D}$ are predicted to be
equilibrium states. Mathematically, this implies the model must
satisfy the equations of state
\begin{equation} 
    \nabla_{\bm{P}}  F\biggr|_{\text{eq}} = {\bf 0}
\end{equation}
and
\begin{equation}
  \nabla_{\bm{\eta}}  F \biggr|_{\text{eq}} = {\bf 0} \; ,
\end{equation}
where we use the notation
\begin{equation}
  \nabla_{\bm{P}}=(\partial_{P_x}, \partial_{P_y}, \partial_{P_z})
\end{equation}
and
\begin{equation}
      \nabla_{\bm{\eta}}=(\partial_{\eta_{xx}}, \partial_{\eta_{yy}},
      \partial_{\eta_{zz}}, \partial_{\eta_{yz}},
      \partial_{\eta_{xz}},\partial_{\eta_{xy}}) \; .
\end{equation}
We thus have a total of 9 equations, 3 for the polarization components
and 6 for the strain components. By using $F$ as given in
Eq.~(\ref{eq:F}), we get
\begin{equation}
\begin{aligned}
\sum_{\beta}^{p(A)} &A_{\beta}(T^{(s)}) \nabla_{\bm{P}}
f_{\beta}(\mathbf{P})\biggr|_{\bm{P}^{(s)}} + \\ +& \sum_{\beta}^{p(B)}
B_{\beta}(T^{(s)}) \nabla_{\bm{P}} f_{\beta}(\mathbf{P}, \bm{\eta})
\biggr|_{\bm{P}^{(s)}, \bm{\eta}^{(s)}} =\bm{\mathcal{E}}^{(s)}
\end{aligned}
\label{eq:eos-P}
\end{equation}
and
\begin{equation}
  \begin{aligned}
\sum_{\beta}^{p(C)} &C_{\beta}(T^{(s)}) \nabla_{\bm{\eta}}f_{\beta}(\bm{\eta})\biggr|_{\bm{\eta}^{(s)}} + \\
+& \sum_{\beta}^{p(B)} B_{\beta}(T^{(s)})
\nabla_{\bm{\eta}}f_{\beta}(\mathbf{P}, \bm{\eta})
\biggr|_{\bm{P}^{(s)}, \bm{\eta}^{(s)}} =  \bm{\sigma}^{(s)} \; ,
\end{aligned}
\label{eq:eos-eta}
\end{equation}
where we explicitly indicate that we can construct such relations for
all $s = 1, ..., n$ states in $\mathcal{D}$. Here we also introduce
$p(A)$, $p(B)$ and $p(C)$ as the number of $A_{\beta}$, $B_{\beta}$
and $C_{\beta}$ parameters, respectively, in the model being
considered. Hence, we have $N = 9n$ linear equations and $p =
p(A)+p(B)+p(C)$ unknowns, which will typically yield an overdetermined
system.

For convenience, we express the system of equations using the matrix
notation
\begin{equation}
  \dbs{M} \bm{\phi} = \bm{y} \; ,
  \label{eq:sys}
\end{equation}
where $\bm{y}$ is a vector of dimension $N$ that contains the
right-hand side of the equations of state, that is, 3 electric field
components and 6 stress components for each of the $n$ states in
$\mathcal{D}$. Then, $\bm{\phi}$ is a vector of dimension $p$ that
contains the $\phi_{\beta}$ coefficients to be fitted. Finally,
$\dbs{M}$ is a matrix of dimensions $N \times p$ that is populated
with the derivatives -- with respect to polarization and strain
components -- of the $f_{\beta}$ invariants.

Such an overdetermined system of equations does not have an exact
solution. Hence, we fit the coefficients $\bm{\phi}$ so as to minimize
the discrepancy between the left- and right-hand sides of
Eq.~(\ref{eq:sys}). To this end, we introduce the error function
\begin{equation}
    \mathsf{E}(\bm{\phi}) = \frac{1}{N}||\dbs{M} \bm{\phi} -
    \bm{y}||^{2} \; .
    \label{eq:E}
\end{equation}
If we minimize this quantity with respect to $\bm{\phi}$, by solving
for $\partial_{\bm{\phi}} \mathsf{E} = 0$, we obtain a solution known
as the normal equation:\cite{press-book2007}
\begin{equation}
    \bm{{\phi}}_{\text{min}} = (\dbs{M}^T\dbs{M})^{-1} \dbs{M}^T
    \bm{y} \; .
    \label{eq:solution}
\end{equation}
Hence, we take advantage of the fact that Landau potentials are linear
in the parameters to be fitted, which makes it possible to estimate
their values by solving a simple polynomial regression. Thus, the
calculation of the parameters of a given Landau model is exact
and very fast.

\subsubsection{Selecting the best Landau model}
\label{sec:select}

We are left with the task of selecting an optimum Landau potential
from the thousands of variants we can easily generate and fit. Our
optimum potential should be as simple as possible (to facilitate our
physical understanding of the dominant effects) while being
sufficiently accurate (i.e., we want a small $\mathsf{E}(\bm{\phi_{\rm
    min}})$) and predictive (i.e., we want our potential to be as
reliable as possible when applied to situations not considered in the
TS).

As it is known, minimizing an error function like our
$\mathsf{E}(\bm{\phi})$ is not a sufficient criterion to decide how
many (and which) couplings to include in a model potential, as this
will typically lead to overfitting and a reduced predictive power of
the resulting overly-complex models. Then, our experience with
cross-validation schemes -- both from this work and previous related
studies\cite{escorihuelasayalero17} -- is not convincing, as we find
that the usual approaches (from the simple ``leave-p-out''
method\cite{hastie-book2009} to consideration of a proper validation set of
data equivalent to those in the TS) yield models that seem
unnecessarily complex and offer poor predictive power.

\begin{figure}
    \centering
    \includegraphics[width=0.8\linewidth]{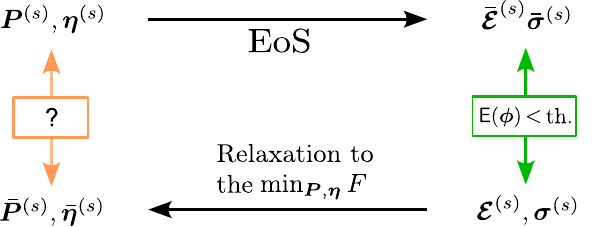}
    \caption{Sketch illustrating our fitting and validation procedures, and how the latter tests the predictive power of the models (see text).}
    \label{fig:validation}
\end{figure}

Here we introduce our own validation scheme, designed to test the
models' predictive power in a way that, while restricted, proves
effective. Figure~\ref{fig:validation} helps illustrate our
approach. Given a model defined by $\bm{\phi}_{\rm min}$, we can use
the information about an equilibrium state,
$\{\bm{P}^{(s)},\bm{\eta}^{(s)};T^{(s)}\}$, to derive the
corresponding external fields from Eqs.~(\ref{eq:eos-P}) and
(\ref{eq:eos-eta}). We express such fields as
\begin{equation}
\bar{\bm{\mathcal E}}^{(s)} = \bm{\mathcal E}[\mathbf{P}^{(s)},
  \bm{\eta}^{(s)};T^{(s)}]
\end{equation}
and
\begin{equation}
\bar{\bm{\sigma}}^{(s)} = \bm{\sigma}[\mathbf{P}^{(s)},
  \bm{\eta}^{(s)};T^{(s)}] \; .
\end{equation}
As indicated in Fig.~\ref{fig:validation}, the model parameters are
computed to minimize the differences between $\{\bm{\mathcal
  E}^{(s)},\bm{\sigma}^{(s)}\}$ and $\{\bar{\bm{\mathcal
    E}}^{(s)},\bar{\bm{\sigma}}^{(s)}\}$. Hence, if our model is
accurate, we can expect these quantities to be close.

Interestingly, because we are dealing with equilibrium states, we can
also go in the opposite direction: given $\{\bm{\mathcal
  E}^{(s)},\bm{\sigma}^{(s)},T^{(s)}\}$, we can run a numerical
simulation to find the minimum of the potential $F({\bm
  P},\bm{\eta};\bm{\mathcal E}^{(s)},\bm{\sigma}^{(s)},T^{(s)})$. Let
us write the solution as
\begin{equation}
    \bar{\bm{P}}^{(s)} = \bm{P}[\bm{\mathcal{E}}^{(s)},
      \bm{\sigma}^{(s)}, T^{(s)}]
\end{equation}
and
\begin{equation}
    \bar{\bm{\eta}}^{(s)} = \bm{P}[\bm{\mathcal{E}}^{(s)},
      \bm{\sigma}^{(s)}, T^{(s)}] \; .
\end{equation}
Noting that the Landau potential will typically present multiple
minima, it is convenient to start the mentioned minimization from
$\{\bm{P}^{(s)},\bm{\eta}^{(s)}\}$, i.e., the actual state in the
TS. Next, we introduce the new error function
\begin{equation}
  \mathsf{e} = \max_{s} \| \bm{P}^{(s)} - \bar{\bm{P}}^{(s)} \| +
  \max_{s} \| \bm{\eta}^{(s)} - \bar{\bm{\eta}}^{(s)} \| \; ,
  \label{eq:e}
\end{equation}
recalling that we are using rescaled dimensionless quantities (more in
Section~\ref{sec:dimensionless}). This error function quantifies the difference between
$\{\bm{P}^{(s)},\bm{\eta}^{(s)}\}$ and
$\{\bar{\bm{P}}^{(s)},\bar{\bm{\eta}}^{(s)}\}$; thus, it allows us to
test the predictive error of a model.

While this may seem a trivial exercise, we find it proves very
effective to detect (and discard) deficient models. For example,
whenever an excessively complex potential presents spurious local
minima, large values of $\mathsf{e}$ are usually
obtained. Also, whenever the potential is very flat -- e.g., close to
the ferroelectric phase transition --, the error function $\mathsf{E}$
has difficulty to capture the local minima, as local maxima also
satisfy the equilibrium condition; in such cases, seemingly sound models
may render small values of $\mathsf{E}(\bm{\phi}_{\rm min})$ but
comparatively large values of $\mathsf{e}$.

Indeed, as shown below, we find that minimizing this new error
function constitutes the best strategy to identify optimum models in a
reliable and automatic manner. Admittedly, the procedure is
computationally heavier than a simple validation step, as it involves
a large number of free energy minimizations. Yet, given the relative
simplicity of our Landau potentials -- compared to, e.g., atomistic
force fields --, it remains affordable using standard desktop
computers. Let us also stress that this validation procedure takes
advantage of a specific feature of the TS we use here, namely, that
all the states in $\mathcal{D}$ are equilibrium configurations that
have a (local) minimum of the free energy associated to them. By
contrast, in the usual MLIP approaches, the TS is composed of
non-equilibrium configurations that are not minima of any energy
function; hence, in principle, an analogous test of the predictive power of a machine-learned interatomic potential would not be possible.

\section{Numerical application}

In this section we justify our choice of PbTiO$_{3}$ as a convenient
material to illustrate our method. Then, we describe how we generate
the dataset $\cal{D}$ and the second-principles atomistic model used
for that purpose. Finally, we specify the way in which we rescale the
quantities entering the Landau potential.

\subsection{Ferroelectric PbTiO$_{3}$}
\label{sec:pto}

We work with ferroelectric perovskite oxide PbTiO$_3$ (PTO) to
illustrate our methodology. PTO is a very relevant material, both
scientifically and technologically, and remains one of the best
studied ferroelectrics.\cite{rabe-book2007} Further, it provides us with a test case that
is both challenging (as regards the subtle physical couplings involved) and
formally simple (because of the high symmetry, $Pm\bar{3}m$, of the
reference $\bm{P}=\bm{0}$ phase).

Experimentally, PTO displays a discontinuous ferroelectric transition,
at T$_{\rm C} = 760$~K, between a cubic paraelectric phase (with space
group $Pm\bar{3}m$) and a tetragonal ferroelectric phase
($P4mm$).\cite{haun87} The spontaneous polarization is accompanied by
a significant lattice deformation, as the unit cell develops a $c/a$
aspect ratio of about 1.07 at room temperature. Accordingly, DFT
studies evidence a strong coupling of polarization and strain.\cite{kingsmith94} For
example, and most remarkably, it has been predicted\cite{wojdel13}
that, if a temperature-independent cubic cell (i.e., $\bm{\eta} =
\bm{0}$) is imposed, PTO undergoes a continuous ferroelectric phase
transition, at a much reduced temperature, to a phase that is
rhombohedral ($R3m$) instead of tetragonal. (In these conditions, PTO
is predicted to undergo an additional transition, at a lower
temperature, to another rhombohedral ($R3c$) structure that resembles
the most common phase of BiFeO$_{3}$.\cite{wojdel13} Treating this
case within a Landau approach would require considering extra order parameters in our model --
i.e., octahedral tilts -- and we do not consider it here.) Let us
stress that these are highly non-trivial behaviors that stem from the
complex multi-minima energy landscape of PTO. Hence, this material
poses a singular challenge to our method for computing free energy
potentials. Indeed, such peculiar conditions (e.g., $\bm{\eta} =
\bm{0}$) will not be part of our TS and, thus, it is an open and
intriguing question whether our potentials will capture the mentioned
effects.

\subsection{Atomistic simulations and training set}
\label{sec:ts}

To generate the training set $\cal{D}$, we employ the
``second-principles'' atomistic effective potential for PTO described
in Ref.~\onlinecite{wojdel13}. This model has been used in numerous
studies of PTO and related compounds (e.g., PTO-based superlattices),
and it has been shown to reproduce the behavior of the material in a
way that is qualitatively and quasi-quantitatively
correct.\cite{wojdel14a,zubko16,goncalves19,das19,graf21} The only noteworthy
discrepancy between the model predictions and experimental results
concerns the quantification of the ferroelectric transition
temperature: for bulk PTO, Monte Carlo simulations using the
second-principles model yield $T_{\rm C} \approx 510$~K, while the
experimental Curie point occurs at 760~K.\cite{wojdel13} While
significant, this disagreement is irrelevant in the current context, as we are not concerned by agreement with experiment. Here our focus is on capturing non-trivial thermodynamic behaviors using simple
Landau-like potentials, and our atomistic model for PTO does capture many such complex behaviors.

It is worth noting here that our second-principles model for PTO involves a relatively low-order expansion of the energy around a cubic reference structure: up to the 4th order for the terms that depend on atomic displacements alone, quadratic in the terms related to strains alone, and with strain-phonon couplings that are linear in strain and quadratic in the atomic displacements.\cite{wojdel13} Such a low-order model is nevertheless capable of reproducing non-trivial effects, such as PTO's weakly first-order phase transition, whose description is known to require a 6th-order free energy potential $\tilde{F}$.\cite{landau-book1980,devonshire54} This is an example of how the free energy landscape for the polarization gets renormalized by thermal fluctuations and by averaging over all non-essential atomic degrees of freedom, yielding behaviors that may not be immediately obvious from the form of the underlying atomic interactions. This point is discussed in some detail in Refs.~\onlinecite{iniguez01} and \onlinecite{wojdel14b}. 

To compute the equilibrium properties of bulk PTO at varying values of
temperature, electric field and stress, we solve our second-principles
model using Monte Carlo (MC) simulations. We typically work with a supercell that is a 
$10\times 10\times 10$ repetitions of the elemental perovskite 5-atom unit, thus corresponding to 5~000 atoms, assuming periodic boundary conditions. Close to the computed
transition temperature, between 460~K and 550~K, we increase the
supercell size to $12\times 12\times 12$, corresponding to 8~640
atoms, to mitigate finite-size effects. For each case considered, we
perform 10~000~MC sweeps for thermalization of the system, followed by
40~000 sweeps to calculate the thermal averages of polarization and
strain. We checked that these choices yield sufficiently converged
results. Typically, at each temperature considered, we first solve the
case where the external fields are set to zero, and use a configuration characteristic of the corresponding equilibrium state as the starting point of the
calculations with fields applied.

We build two training sets, $\tilde{\mathcal D}$ and ${\mathcal D}$,
to train the $\tilde{F}({\bm P};\bm{\mathcal E},T)$ and $F({\bm
  P},\bm{\eta};\bm{\mathcal{E}},\bm{\sigma},T)$ models,
respectively. To build $\tilde{\mathcal D}$ we consider 21 different
temperatures ranging from 50~K to 1000~K, with a higher concentration
of temperatures close to $T_{\rm C}$. For each temperature, we
consider 27 cases with an electric field applied (and zero applied stress);
the field components are chosen randomly with a maximum magnitude of
100~MV/m. All the states in $\tilde{\mathcal D}$ are obtained in zero-stress conditions. 
For the second training set, ${\mathcal D}$, we consider 7
temperatures ranging from 100~K to 700~K. For each temperature we
include the corresponding zero-stress states contained in
$\tilde{\mathcal D}$. Additionally, for each temperature we include 12
cases with applied external stress (and zero applied electric field); the
stress components are randomly chosen with a maximum magnitude of
1\;GPa. All in all, $\tilde{\mathcal{D}}$ contains 567 entries
corresponding to equilibrium states at zero stress, yielding 1~701
equilibrium relations (Eq.~(\ref{eq:eos-P})). Meanwhile, ${\mathcal{D}}$ consists
of 273 entries, yielding 2~457 relations (Eqs.~(\ref{eq:eos-P}) and
(\ref{eq:eos-eta})).

\subsection{Dimensionless quantities}
\label{sec:dimensionless}

It is convenient to use natural scales for the various physical
variables involved in the problem, so as to work with dimensionless
magnitudes. This allows us to combine the errors associated to
different quantities, as e.g. done in Eqs.~(\ref{eq:E}) and
(\ref{eq:e}) for electric and elastic variables.

We thus rescale energy and polarization so that the tetragonal ground
state of our simulated bulk PTO (in the limit of 0~K) has an energy
$\Delta F = F - F_{0} = -1$ and a polarization $P_{z} = 1$. We further
rescale the temperatures so that $T_{\rm C} = 1$. As for the strains,
they are dimensionless by definition and we find it appropriate to not
rescale them at all. The actual values of energy,
polarization, and temperatures can be recovered by simply noting that
the unscaled quantities are: $\Delta F = -4.64$~GJ\;m$^{-3}$, $P_{z} =
0.992$~C\;m$^{-2}$ and $T_{\rm C} = 510$~K.

As discussed below, we checked these choices and found it unnecessary
to further weight the electric and elastic contributions to our error
functions. Thus, the scalings just described enable a balanced fit
that treats well all the variables in the problem. Note that this
conclusion might be specific of the problem studied in this work;
the scaling of the variables in the model should be considered
on a case-by-case basis. As a default strategy, one could use the
corresponding standard deviation in the TS data as a natural scale for
each variable involved in the free energy model. Such a choice seems
to render robust results in MLIP training schemes that combine errors
in energy, forces and stresses.\cite{jinnouchi19}

\section{Results}

We discuss first Landau potentials that are functions of the
polarization alone, then potentials where strains are explicitly
treated.

\subsection{$\tilde{F}(\mathbf{P}; \bm{\mathcal{E}}, T)$ models}

Since these models do not treat strain explicitly, they cannot
reproduce the dependence of the system's properties as a function of
stress. Hence, to fit $\tilde{F}(\mathbf{P}; \bm{\mathcal{E}}, T)$
models, we use the zero-stress data in $\tilde{\mathcal D}$ defined
above. Similarly, the error functions $\tilde{\mathsf{E}}$ and
$\tilde{\mathsf{e}}$ are obtained from Eqs.~(\ref{eq:E}) and
(\ref{eq:e}), respectively, by leaving out the contributions from
stress and strain.

\begin{figure}
    \centering
    \includegraphics[width=0.85\linewidth]{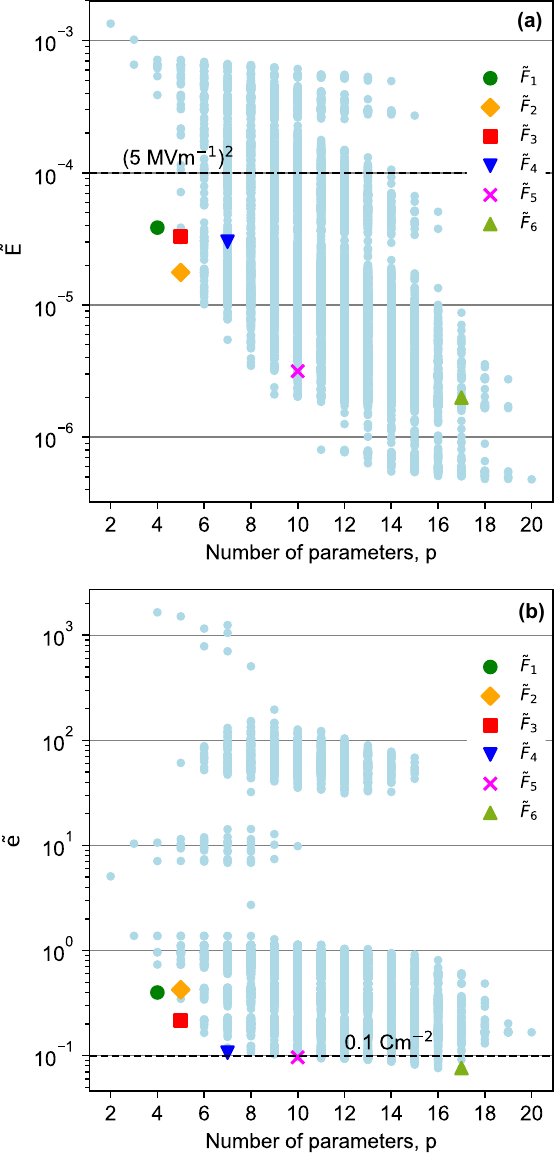}
    \caption{(a) Error function $\tilde{\mathsf{E}}$ as a function of the number of parameters in the model ($p$). 
(b) Predictive error $\tilde{\mathsf{e}}$ as a function of $p$. Coloured symbols highlight the selected models, from $F_{1}$ to $F_6$, discussed in the text. Error thresholds mentioned in the text are indicated with dashed lines.}
    \label{fig:P-models}
\end{figure}

As described in Section~\ref{sec:approach}, our procedure consists in
fitting all possible models up to the 8th order; that is, we consider
all possible combinations of the invariants $f_{\beta}(\bm{P})$ in
Table~\ref{tab:F}, albeit the mentioned simplifications (e.g., the
$\tilde{A}_{2i}^{(0)}$ and $\tilde{A}_{2i}^{(1)}$ parameters are always included). Our results are
summarized in Fig.~\ref{fig:P-models}(a), from which we can draw two
main conclusions. On the one hand, the average error
$\tilde{\mathsf{E}}$ tends to decrease as we increase the number of
parameters in the model, but with a large spread for models with the
same number of couplings. On the other hand, we obtain remarkably
accurate models with a relatively small number of parameters.

For the sake of the argument, let us consider a model to be
  reasonably accurate if $\tilde{\mathsf{E}} < 10^{-4}$. This
threshold, marked with a dashed line in Fig.~\ref{fig:P-models}(a),
corresponds to having electric fields reproduced to within
5~MV~m$^{-1}$ -- i.e., 1/20~times the largest applied field considered
in our TS --, which would seem sufficient for most use cases. (Having
thermodynamic properties predicted to within a 5~\% is, at
least, on par with the accuracy of present first-principles methods.)
Then, models with as few as 4 parameters already give an acceptable
description of our data. For example, this is the case of the model
labeled $\tilde{F}_{1}$ in Fig.~\ref{fig:P-models}(a), which is given by
\begin{equation}
\begin{split}
    \tilde{F}_{1} = & \;(\tilde{A}_{2i}^{(0)} +
    \tilde{A}_{2i}^{(1)}T) (P_{x}^{2} + P_{y}^{2} + P_{z}^{2}) \\ &
    \;+ \tilde{A}_{22}^{(0)}(P_{x}^{2} P_{y}^{2} + P_{x}^{2} P_{z}^{2}
    + P_{y}^{2} P_{z}^{2}) \\ & \;+ \tilde{A}_{6i}^{(0)}(P_{x}^{2} +
    P_{y}^{2} + P_{z}^{2})^{3} \; ,
\end{split}
\end{equation}
and for which we report the fitted parameters in
Table~\ref{tab:P-models}. The accuracy of this simple model in
reproducing $\tilde{\mathcal D}$ can be better appreciated in Suppl.
Fig.~1. This figure also shows the result for the best 5-parameter
model identified by our method, labeled $\tilde{F}_{2}$ in
Fig.~\ref{fig:P-models}(a), which includes the same couplings as
$\tilde{F}_{1}$ plus the $\tilde{A}_{42}^{(0)}$ term (see
Table~\ref{tab:P-models}). Hence, if we take the error function
$\tilde{\mathsf{E}}$ as our sole quality criterion, models
$\tilde{F}_{1}$ and $\tilde{F}_{2}$ would seem sufficient choices to
describe PTO.

\begin{figure}
    \includegraphics[width=0.8\linewidth]{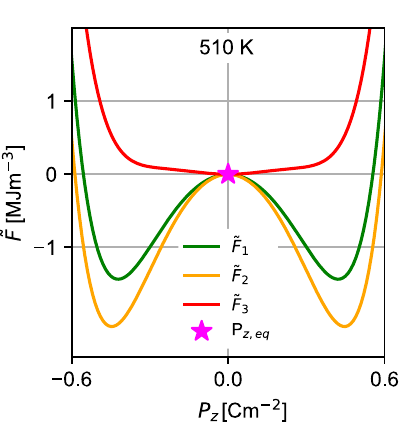}
    \centering
    \caption{Free energy landscape as a function of $P_{z}$ (for $P_{x}=P_{y}=0$) at $T=510$ K and $\bm{\mathcal{E}}= \bm{0}$ for models $\tilde{F}_1$, $\tilde{F}_2$, and $\tilde{F}_3$ (see text). The magenta star indicates the equilibrium polarization $P_{z,eq}$ present in the TS. Note that the free energy and polarization are given in SI units.}
    \label{fig:landscape}
\end{figure}

However, as mentioned in Section~\ref{sec:approach}, minimizing
$\tilde{\mathsf{E}}$ does not guarantee the predictive power of our
models, as illustrated by the following example. Let us consider three
models, the already mentioned $\tilde{F}_{1}$ and $\tilde{F}_{2}$ as
well as an additional 5-parameter model labeled $\tilde{F}_{3}$ in
Fig.~\ref{fig:P-models}(a). Figure~\ref{fig:landscape} shows the free
energy as a function of $P_{z}$ (and for $P_{x}=P_{y}=0$) given by
these three models at $T=510$~K, right above the ferroelectric
transitions temperature predicted by our atomistic simulations. We
know that, at this temperature and for zero applied electric field,
the equilibrium polarization of the material is zero, i.e., the Landau
potential should have a minimum at the origin. However, models
$\tilde{F}_{1}$ and $\tilde{F}_{2}$ render a double well potential,
with a maximum at ${\bf P}={\bf 0}$, a qualitatively incorrect
result. By contrast, model $\tilde{F}_{3}$ -- which has a bigger error
$\tilde{\mathsf{E}}$ than model $\tilde{F}_{2}$ -- predicts a very
flat energy landscape with a minimum at the origin. Indeed, as regards
predicting the behavior of the material near $T_{\rm C}$ in a
qualitatively correct way, $\tilde{F}_{3}$ clearly outperforms
$\tilde{F}_{2}$. Next we show how we can identify and select such
reliable models automatically.

\begin{table*}
\caption{Parameters for selected $\tilde{F}({\bm P};\bm{\mathcal
    E},T)$ models given in SI units.}  \centering
\begin{tabular}{c c c S S S S S S}
  \hline\hline
  \addlinespace[2ex]
  & Parameter & Units & \multicolumn{1}{c}{$\tilde{F}_1$} & \multicolumn{1}{c}{$\tilde{F}_2$} & \multicolumn{1}{c}{$\tilde{F}_3$} & \multicolumn{1}{c}{$\tilde{F}_4$} &\multicolumn{1}{c}{$\tilde{F}_5$} & \multicolumn{1}{c}{$\tilde{F}_6$} \\ \hline 
  \addlinespace[1ex]
    \multirow{10}{*}{\rotatebox{90}{$T$-independent terms}}
 &$\tilde{A}_{2i}^{(0)}$& 10$^{8}$ \;C$^{-2}$\;m$^{2}$\;N & -3.94679 & -4.02122 & -3.75628 & -3.63736 & -3.20683 & -3.35136 \\ 
 &$\tilde{A}_{4i}^{(0)}$ &10$^{8}$ \;C$^{-4}$\;m$^{6}$\;N &  &  & -0.27596 & -0.56378 & -1.97729 & -1.17002 \\ 
 &$\tilde{A}_{22}^{(0)}$ &10$^{8}$ \;C$^{-4}$\;m$^{6}$\;N &  2.92026 &  4.22073 &  2.92379 &  2.73622 &  3.19689 &  1.59453 \\ 
 &$\tilde{A}_{6i}^{(0)}$ &10$^{8}$ \;C$^{-6}$\;m$^{10}$\;N &  1.34097 &  1.38310 &  1.47317 &  1.62257 &  2.46231 &  1.40375 \\ 
 &$\tilde{A}_{42}^{(0)}$ &10$^{8}$ \;C$^{-6}$\;m$^{10}$\;N &  & -1.98136 &  &  &  1.50262 &  4.01424 \\ 
 &$\tilde{A}_{222}^{(0)}$& 10$^{8}$ \;C$^{-6}$\;m$^{10}$\;N &  &  &  &  7.76192 &  &  15.87736 \\ 
 &$\tilde{A}_{8i}^{(0)}$& 10$^{8}$ \;C$^{-8}$\;m$^{14}$\;N &  &  &  &  &  &  0.42261 \\ 
 &$\tilde{A}_{422}^{(0)}$& 10$^{8}$ \;C$^{-8}$\;m$^{14}$\;N &  &  &  &  &  & -16.10410 \\ 
 &$\tilde{A}_{44}^{(0)}$& 10$^{8}$ \;C$^{-8}$\;m$^{14}$\;N &  &  &  &  & -5.02926 & -4.20465 \\ 
 &$\tilde{A}_{62}^{(0)}$& 10$^{8}$ \;C$^{-8}$\;m$^{14}$\;N & & & & & -2.66843 & -3.80982 \\  \addlinespace[1ex]  \hline \addlinespace[1ex] 
     \multirow{7}{*}{\rotatebox{90}{$T$-dependent terms}}

 &$\tilde{A}_{2i}^{(1)}$& 10$^{6}$ \;C$^{-2}$\;m$^{2}$\;N\;K$^{-1}$ &  0.74973 &  0.75719 &  0.74130 &  0.72655 &  0.65142 &  0.66895 \\ 
 &$\tilde{A}_{4i}^{(1)}$ &10$^{6}$ \;C$^{-4}$\;m$^{6}$\;N\;K$^{-1}$ &  &  &  &  0.02726 &  0.25120 &  0.15726 \\ 
 &$\tilde{A}_{22}^{(1)}$ &10$^{6}$ \;C$^{-4}$\;m$^{6}$\;N\;K$^{-1}$ &  &  &  &  &  &  0.15596 \\ 
 &$\tilde{A}_{6i}^{(1)}$& 10$^{6}$ \;C$^{-6}$\;m$^{10}$\;N\;K$^{-1}$ &  & &  &  & -0.09493 & -0.02123 \\ 
 &$\tilde{A}_{222}^{(1)}$& 10$^{6}$ \;C$^{-6}$\;m$^{10}$\;N\;K$^{-1}$ &  &  &  &  &  &  0.23905 \\ 
 &$\tilde{A}_{422}^{(1)}$ &10$^{6}$ \;C$^{-8}$\;m$^{14}$\;N\;K$^{-1}$ &  &  &  &  &  & -2.36708 \\ 
 &$\tilde{A}_{44}^{(1)}$& 10$^{6}$ \;C$^{-8}$\;m$^{14}$\;N\;K$^{-1}$ & &  &  &  & & -1.21606 \\ 
\addlinespace[1ex]
\hline\hline
\end{tabular}
\label{tab:P-models}
\end{table*}

To do this, we use the second error function, $\tilde{\mathsf{e}}$
(Eq.~(\ref{eq:e})), introduced in Section~\ref{sec:select}, which
quantifies the models' ability to predict the equilibrium states of
the TS by energy minimization. Our results are summarized in
Fig.~\ref{fig:P-models}(b). According to this new criterion, $\tilde{F}_{3}$ is
the best choice of 5-parameter model, while the performance of $\tilde{F}_{2}$
is relatively modest. If we set $\tilde{\mathsf{e}} < 10^{-1}$ as the
threshold for accuracy -- which corresponds to a maximum error of
about 0.1~C~m$^{-2}$ for the predicted polarization --, we find that the
model labeled $\tilde{F}_{4}$ (10 parameters, see Table~\ref{tab:P-models}) is
the simplest model that complies with our requirements.

It is worth noting that the predictive error $\tilde{\mathsf{e}}$
allows us to identify the level of complexity beyond which our
models do not improve. Indeed, Fig.~\ref{fig:P-models}(b) includes
results for 156 different models of 17 parameters, as well as 55
different models of more than 17 parameters. We find that, beyond 17 independent couplings, $\tilde{\mathsf{e}}$ starts to grow with the number of parameters, the absolute best model being the one labeled $\tilde{F}_{6}$. Indeed, models with more than 17 parameters continue to improve their accuracy, as shown in
Fig.~\ref{fig:P-models}(a), but yield increasingly worse predictions, as shown in Fig.~\ref{fig:P-models}(b).

\begin{figure}
    \centering
    \includegraphics[width=0.8\linewidth]{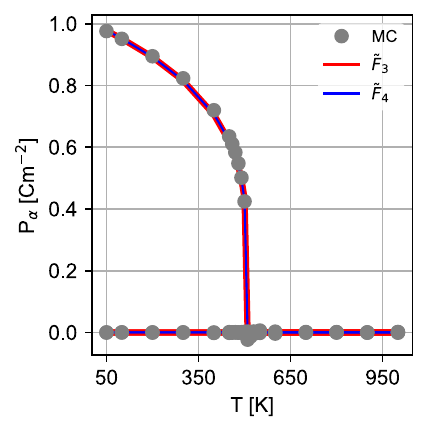}
    \caption{Equilibrium polarization as a function of temperature with $\bm{\mathcal{E}} = \bm{0}$. The grey data points represent values obtained from second-principles Monte Carlo simulations and are part of our dataset, while the red and blue curves correspond to the predictions of models $\tilde{F}_3$ and $\tilde{F}_4$, respectively. Note that the polarization and temperature are given in SI units.}
    \label{fig:P-vs-T}
\end{figure}

The prediction error $\tilde{\mathsf{e}}$ quantifies the maximum
discrepancy across the entire dataset. As mentioned above, cases where
the free energy surface is very flat -- most notably, close to $T_{\rm
  C}$ -- can be expected to be particularly challenging. This can be
appreciated in Supp. Fig.~2, where we show the maximum prediction
error as a function of temperature. By examining these detailed
results, we can finetune the choice of Landau model according to our
needs in particular applications. For example, if we are interested in
describing PTO in a broad temperature range, and if it is sufficient
to have a qualitatively correct description of the transition region,
$\tilde{F}_{3}$ might be our optimal choice. This model is given by
\begin{equation}
    \begin{split}
    \tilde{F}_{3} = & \;(\tilde{A}_{2i}^{(0)} +
    \tilde{A}_{2i}^{(1)}T) (P_x^2 + P_y^2 + P_z^2) \\ &\;+
    \tilde{A}_{4i}^{(0)}(P_x^2 + P_y^2 + P_z^2)^2 \\ &\;+
    \tilde{A}_{22}^{(0)}(P_x^2 P_y^2 + P_x^2 P_z^2 + P_y^2 P_z^2)
    \\ &\;+ \tilde{A}_{6i}^{(0)}(P_x^2 + P_y^2 + P_z^2)^3,
    \end{split}
    \label{eq:f3}
\end{equation}
with parameter values in Table~\ref{tab:P-models}. Figure~\ref{fig:P-vs-T} shows the temperature dependence of the polarization of PTO as obtained from $\tilde{F}_{3}$ and its comparison with second-principles data; the agreement is essentially perfect. The agreement remains perfect for model $\tilde{F}_{4}$.

By inspection of Table~\ref{tab:P-models}, one can note that the
best-performing models highlighted in this section share the majority
of their couplings. This further supports the robustness of our
approach and the physical meaningfulness of the interaction terms identified
by our automatic ML procedure.

\begin{figure}
    \centering
    \includegraphics[width=0.85\linewidth]{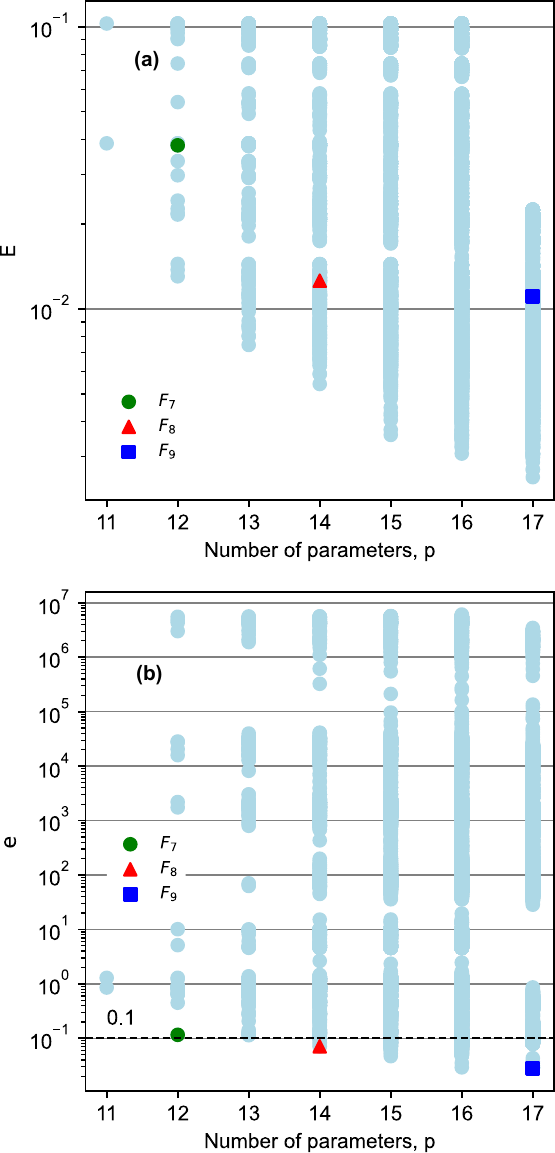}
       \caption{(a) Error function $\mathsf{E}$ as a function of the polynomial degree $p$.
(b) Predictive error $\mathsf{e}$ as a function of $p$. The dashed line indicates the $0.1$ [arb. unit.] threshold.
Coloured symbols highlight the selected models ($F_7$, $F_{8}$, and $F_9$) as shown in the legend.}
    \label{fig:Peta-models}
\end{figure}

\subsection{$F({\bm P},\bm{\eta};\bm{\mathcal E},\bm{\sigma},T)$ models}
\label{sub: P-ETA_results}

We now consider models including strain, which we fit using the
$\mathcal{D}$ training set described in Section~\ref{sec:ts}. Figure~\ref{fig:Peta-models} summarizes our
results, i.e., the errors obtained from a direct minimization of
$\mathsf{E}$ (a) and the validation error $\mathsf{e}$ (b). We focus
on models containing between 11 and 17 parameters, which we find are reasonably
accurate.

As in the case of the $\tilde{F}({\bm P};\bm{\mathcal E},T)$ models,
we find that the most accurate potentials (smallest $\mathsf{E}$) are
not the ones delivering most robust predictions (smallest
$\mathsf{e}$). As in the previous section, we establish a threshold of
$\mathsf{e} < 10^{-1}$ for the acceptable model accuracy, which
corresponds to a maximum error of about 0.1~C~m$^{-2}$ for the
predicted polarization and about 0.4\% for the predicted strain. We
find that the 14-parameter model labeled $F_{7}$ gives us an
acceptable description of the system. We also see that, if we extend
our analysis up to models with 17 parameters, we can keep improving
the predictive error, models labeled $F_{8}$ and $F_{9}$ being two
representative examples. Supplementary Figure~3 includes more details
on how these models reproduce the TS data, and the mentioned models
are fully specified in Table~\ref{tab:PETAmodels}.

\begin{table*}
\caption{Parameters of selected $F({\bm P}, {\bm \eta};\bm{\mathcal
    E}, {\bm \sigma},T)$ models given in SI units.}
\begin{tabular}{cccSSS}
  \hline\hline
  \addlinespace[2ex]
  &Parameter & Units &\multicolumn{1}{c}{$F_7$} & \multicolumn{1}{c}{$F_8$} & \multicolumn{1}{c}{$F_9$} \\ \hline 
  \addlinespace[1ex]
  \multirow{13}{*}{\rotatebox{90}{$T$-independent terms}}
&$A_{2i}^{(0)}$& 10$^{8}$ \;C$^{-2}$\;m$^{2}$\;N & -3.81307 & -3.17357 & -3.11360 \\ 
&$A_{4i}^{(0)}$& 10$^{8}$ \;C$^{-4}$\;m$^{6}$\;N &  0.90945 &  0.56767 & -1.62015 \\ 
&$A_{22}^{(0)}$& 10$^{8}$ \;C$^{-4}$\;m$^{6}$\;N & -1.67506 & -5.00439 & -1.63859 \\ 
&$A_{6i}^{(0)}$& 10$^{8}$ \;C$^{-6}$\;m$^{10}$\;N &  2.01344 &  2.58481 &  4.70951 \\ 
&$A_{42}^{(0)}$& 10$^{8}$ \;C$^{-6}$\;m$^{10}$\;N &  &  & -1.01874 \\ 
&$A_{62}^{(0)}$& 10$^{8}$ \;C$^{-8}$\;m$^{14}$\;N &  &  & -4.23945 \\ 
&$B_{12}^{(0)}$&  10$^{8}$ \;C$^{-2}$\;m$^{2}$\;N & -69.23618 & -99.33400 & -70.24195 \\ 
&$B_{14}^{(0)}$& 10$^{8}$ \;C$^{-4}$\;m$^{6}$\;N &  &  & -31.37833 \\ 
&$B_{1^{'},11}^{(0)}$& 10$^{8}$ \;C$^{-2}$\;m$^{2}$\;N &  0.01283 &  0.01284 &  0.01283 \\ 
&$C_1^{(0)}$& 10$^{11}$ \;m$^{-2}$\;N & -0.00068 &  0.00382 &  0.00293 \\ 
&$C_2^{(0)}$& 10$^{11}$ \;m$^{-2}$\;N &  0.81154 &  0.99169 &  1.04524 \\ 
&$C_{11}^{(0)}$& 10$^{11}$ \;m$^{-2}$\;N &  0.57923 &  0.61994 &  0.67598 \\ 
&$C_{2^{'}}^{(0)}$& 10$^{11}$ \;m$^{-2}$\;N &  0.50354 &  0.50354 &  0.50354  \\ \addlinespace[1ex]  \hline    \addlinespace[1ex]
\multirow{6}{*}{\rotatebox{90}{$T$-depend. terms}}
&$A_{2i}^{(1)}$& 10$^{6}$ \;C$^{-2}$\;m$^{2}$\;N\;K$^{-1}$ &  0.83591 &  0.74608 &  0.72419 \\ 
&$A_{4i}^{(1)}$& 10$^{6}$ \;C$^{-4}$\;m$^{6}$\;N\;K$^{-1}$ &  &  &  0.38654 \\ 
&$A_{22}^{(1)}$& 10$^{6}$ \;C$^{-4}$\;m$^{6}$\;N\;K$^{-1}$ &  &  0.75520 &  \\ 
&$A_{6i}^{(1)}$& 10$^{6}$ \;C$^{-6}$\;m$^{10}$\;N\;K$^{-1}$ &  &  & -0.14247 \\ 
&$B_{12}^{(1)}$&  10$^{6}$ \;C$^{-2}$\;m$^{2}$\;N\;K$^{-1}$ &  &  3.86652 &  \\ 
&$C_1^{(1)}$& 10$^{6}$ \;m$^{-2}$\;N\;K$^{-1}$ & -2.50047 & -3.68843 & -3.78180 \\
\addlinespace[1ex]
\hline\hline
\end{tabular}
\label{tab:PETAmodels}
\end{table*}

Moving beyond 17 parameters involves consideration of a much larger
number of models, and we did not pursue this systematically. Nevertheless, based on a partial
exploration -- by extending models with small $\mathsf{E}$ and
low $\mathsf{e}$ errors --, we obtain evidence suggesting that the predictive error $\mathsf{e}$ is
minimized for a model with 30 parameters.

If we inspect the electric and elastic contributions to $\mathsf{E}$,
it is apparent that the error is dominated by the elastic component
(see Supplementary Note~1 and Supplementary Figure~4). In view of
this, we experimented with reweighting the different contributions to
$\mathsf{E}$ and $\mathsf{e}$, but did not observe any significant
change in the fitted models.

We also considered models with additional (higher order) strain, and found that our automatic procedure continues to select the same -- simple and reasonably accurate -- optimum potentials. Hence, we conclude that the way in which we combine the electric and elastic contributions to the error is sufficient to obtain reliable results.

If we turn to our optimum $F({\bm P},\bm{\eta};\bm{\mathcal
  E},\bm{\sigma},T)$ models, we find that $F_{7}$ has the following
form:
\begin{equation}
    \begin{split}
     F_{7} =  &\;(A_{2i}^{(0)} + A_{2i}^{(1)}T) (P_x^2 + P_y^2 +
     P_z^2) \\ &\;+ A_{4i}^{(0)}(P_x^2 + P_y^2 + P_z^2)^2 \\ &\;+
     A_{22}^{(0)}(P_x^2 P_y^2 + P_x^2 P_z^2 + P_y^2 P_z^2) \\ &\;+
     A_{6i}^{(0)}(P_x^2 + P_y^2 + P_z^2)^3 \\ &\;+
     B_{12}^{(0)} (\eta_{xx} P_{x}^{2} + \eta_{yy}P_{y}^{2} + \eta_{zz}P_{z}^{2}) \\ &\;+
     B_{1^{'}11}^{(0)} (\eta_{yz}P_{y}P_{z} + \eta_{xz}P_{z}P_{x} + \eta_{xy}P_{x}P_{y})\\ &\;+
     (C_1^{(0)} + C_1^{(1)}T) (\eta_{xx} + \eta_{yy} + \eta_{zz}) \\ &\;+
     C_2^{(0)}(\eta_{xx}^2 + \eta_{yy}^2 + \eta_{zz}^2) \\ &\;+
     C_{11}^{(0)} (\eta_{xx} \eta_{zz}+ \eta_{yy} \eta_{xx} + \eta_{yy} \eta_{zz})\\ &\;+
     C_{2^{'}}^{(0)} (\eta_{yz}^{2} + \eta_{xz}^{2}+ \eta_{xy}^{2}),
    \end{split}
\end{equation}
This model is equivalent to $\tilde{F}_{3}$ (Eq.~(\ref{eq:f3})) in what regards the couplings involving the polarization alone, the temperature dependence being restricted to the quadratic parameter $A_{2i}$. Additionally, it contains the lowest-order elastic and polarization-strain couplings, resembling closely (and validating) the simple Landau-like potentials used in the literature.\cite{kingsmith94} It is also worth noting that this model includes, by construction, a term $C^{(1)}_{1}$ that accounts for a variation of the free energy that is linear on both strain and temperature. In essence, this term accounts for thermal expansion, which is of course present in our second-principles simulations. Interestingly, this obviously important factor has never been considered explicitly, as far as we can tell, in phenomenological Landau-type treatments of ferroelectrics.

\begin{figure}
    \centering
    \includegraphics[width=0.8\linewidth]{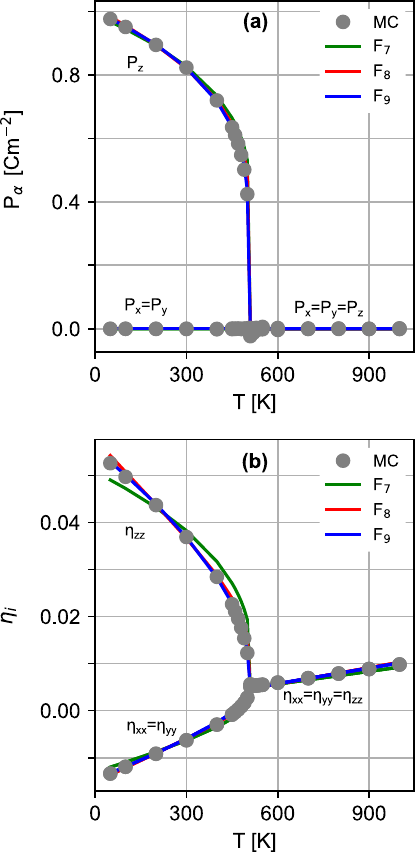}
     \caption{Temperature dependence of polarization (a) and strain (b) components for the models indicated in the legend, which are compared against the second-principles Monte Carlo results in our TS (grey circles). Note that the polarization and temperature are given in SI units.}
    \label{fig:P-eta-vs-T}
\end{figure}

Figure~\ref{fig:P-eta-vs-T} shows the temperature dependence of
polarization and strain obtained by minimizing $F_{7}$, along with a
comparison to the second-principles MC data. The agreement between the
two datasets is essentially perfect, except for a noticeable deviation
for the strain $\eta_{zz}$ parallel to the polarization $P_{z}$. As
for the further improved $F_{8}$ and $F_{9}$, as shown in
Table~\ref{tab:PETAmodels}, they constitute an extension of $F_{7}$
and, most interestingly, they feature additional temperature-dependent
parameters. In this case, as shown in Fig.~\ref{fig:P-eta-vs-T}, we
obtain a perfect agreement between the TS and predicted data for the
temperature-dependent order parameters at zero applied fields.

\section{Discussion}

Having defined and applied our approach to compute optimal
free energy potentials from thermodynamic data, now we discuss some
additional aspects of our work and the opportunities it opens.

\subsection{Non-trivial physical insights and predictions}

In the burgeoning field of MLIPs, some groups are currently focusing
on developing approaches that yield ``interpretable''
potentials.\cite{xie23,jacobs25} Such efforts typically start by introducing a
specific form of the potential -- whose accuracy can be systematically
improved, ideally -- that is tailored to characterize the dominant
interatomic interactions in terms of a relatively small number of
parameters. Usually, such approaches take advantage of decades of work
on effective models and force fields, and somehow constitute a
culmination of those, powered by modern ML tools. We tend to include
in this category diverse approaches ranging from those based on
kernel\cite{jinnouchi19} and many-body potentials\cite{xie23} to the
so-called ``second-principles''\cite{wojdel13,escorihuelasayalero17}
and ``effective Hamiltonian''\cite{zhong95a,ma25} methods based on a
polynomial expansion of the energy around a reference
structure. Beyond their physical transparency (as compared to a
generic neural network at least), such approaches enable a flexible
compromise between the accuracy and the computational cost of the
simulations. Further, we would argue that, the resulting models being
comparatively simple, they can be expected to yield valuable
predictions beyond the TS used to fit them. We have evidence of this
from more than a decade of second-principles work on ferroelectric
oxides,\cite{wojdel13,wojdel14a,zubko16,goncalves19} and we are
observing a similar pattern in ongoing projects where we use
minimalist kernel-based MLIPs to investigate such structurally complex
compounds.\cite{robredo25}

Our approach to compute free energy potentials is physics based and yields overly simple optimal models. Thus, the question naturally emerges: should we expect these models to predict non-trivial physical effects that are not present in the TS data? To explore this matter, we recall the non-trivial behavior of PTO in what concerns the role of strain, as briefly summarized in Section~\ref{sec:pto}. Let us consider such subtle couplings and check whether they are captured by our models. For the sake of concreteness, let us focus on the $\tilde{F}_{4}$ and $F_{7}$ potentials introduced above.

It has been long known from DFT simulations\cite{kingsmith94,wojdel13} that, if we constrain PTO to have the lattice vectors of the extrapolated cubic reference structure (which amounts to choosing $\bm{\eta} = \bm{0}$ in our description), this material displays a rhombohedral ground state with $P_{x}=P_{y}=P_{z}\neq 0$, rather than the usual tetragonal phase. Interestingly, PTO's tendency to display a tetragonal or rhombohedral state is governed by coupling terms like $\tilde{A}_{22}$ or $A_{22}$, which quantify, to the lowest order, the anisotropy of the energy landscape as a function of the polarization orientation. As can be seen from Table~\ref{tab:P-models}, our $\tilde{F}_{4}$ model presents $\tilde{A}_{22} \approx 3 \times 10^8 \;\text{SI}\; > 0$, where the positive value of this parameter implies that a tetragonal state with $\bm{P} \parallel \langle 001 \rangle$ is favored over, e.g., rhombohedral variants with $\bm{P} \parallel \langle 111 \rangle$. By contrast, for $F_{7}$, Table~\ref{tab:PETAmodels} shows that the corresponding parameter is $A_{22} \approx -2 \times 10^8 \;\text{SI}\; < 0$; hence, according to this model, the occurrence of the polarization by itself -- in the absence of strain -- would yield a rhombohedral ground state for PTO. Indeed, in the case of $F_{7}$, we need to allow for polarization and strain to occur simultaneously in order to recover the well-known tetragonal ground state, as indeed shown in Fig.~\ref{fig:P-eta-vs-T}. We thus find that our free energy models capture the subtle polarization-strain couplings that determine the symmetry of the ferroelectric phase of PTO.\cite{kingsmith94,wojdel13} (See Ref.~\onlinecite{kingsmith94} for a deeper discussion on how the strain renormalizes the polarization energy landscape.)

Further, it is widely accepted that in ferroelectric perovskites like PbTiO$_{3}$ or BaTiO$_{3}$ the first-order character of the transition relies on the strain-polarization couplings.\cite{zhong95a,waghmare97} It is also well-known that in simple Landau models -- as those obtained from our optimization procedure -- the character of the transition, continuous or discontinuous, depends on the sign of the $\tilde{A}_{4i}$ parameter, the transition being discontinuous when $\tilde{A}_{4i}< 0$.\cite{landau-book1980,devonshire54} Thus, it is no surprise to find, in Table~\ref{tab:P-models}, that we obtain $\tilde{A}_{4i} \approx -0.6 \times 10^8 \;\text{SI}\; < 0$ for model $\tilde{F}_{4}$. As for model $F_{7}$, Table~\ref{tab:P-models} shows that we obtain $A_{4i} \approx 0.9 \times 10^8 \;\text{SI}\; > 0$, suggesting that PTO should display a continuous ferroelectric phase transition if the strain is kept null. Thus, our simple models capture the polarization-strain couplings that control the nature -- continuous {\it vs}. discontinuous -- of PTO's ferroelectric phase transition.

\begin{figure}
    \centering
    \includegraphics[width=0.8\linewidth]{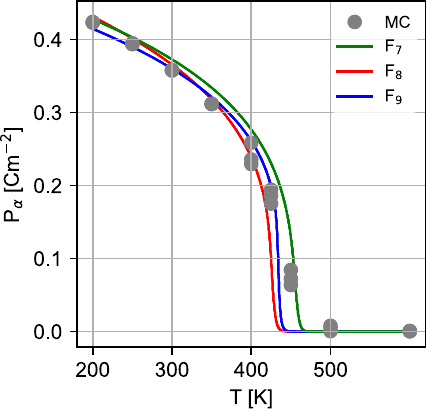}
    \caption{Temperature dependence of the polarization components compared with the predictions of models $F_7$ (green), $F_8$ (red), and $F_9$ (blue) under the elastic condition $\bm{\eta}=\bm{0}$. Black markers represent second-principle data. Note that the polarization and temperature are given in SI units.}
    \label{fig:c2r}
\end{figure}

Having established that our optimal models $\tilde{F}_{4}$ and $F_{7}$ capture well these expected behaviors at a qualitative level, one may wonder to what extent they are also quantitatively accurate. To test this, we perform second-principles Monte Carlo simulations of PTO under the $\bm{\eta} = \bm{0}$ elastic constraint. We obtain a phase transition at about 450~K, from cubic paraelectric to rhombohedral ferroelectric, marked by the spontaneous occurrence of a polarization along a $\langle 111 \rangle$ pseudocubic direction. Additionally, we consider the $F_{7}$, $F_{8}$, and $F_{9}$ models for $\bm{\eta}=\bm{0}$, and compute the minimum of the free energy as a function of temperature. The results from these calculations, shown in Fig.~\ref{fig:c2r}, display a remarkable agreement with the second-principles data. For example, the $F_{7}$ model is able to reproduce the continuous character of the transformation, and it also yields a transition temperature around 450~K. Thus, we find that $F_{7}$ predicts the system's behavior for $\bm{\eta} = \bm{0}$ very well, even at a quantitative level.

We thus find that our optimal models, $\tilde{F}_{4}$ and $F_{7}$, capture all the main physical effects associated to the strain-polarization coupling in a qualitatively correct way that is also physically transparent. Further, the $F_{7}$ model gives a fair quantitative description of our simulated PTO under the $\bm{\eta}=\bm{0}$ elastic constraint. This is a remarkable result, because the TS used to fit this model does not contain any information of the behavior of PTO in states satisfying, not even approximately, the $\bm{\eta}=\bm{0}$ condition (we checked that the configurations in ${\mathcal D}$ present strains with a magnitude of at least 0.5~\%). This agreement strongly supports the choice of polynomial Landau-like models to capture, within ML schemes, the free energy potential of materials displaying non-reconstructive (typically, soft-mode driven) phase transitions, of which PTO is a representative example. More generally, these findings underscore the utility of physics-informed ML approaches and the remarkable predictive power such models can exhibit.

\subsection{Peculiarities of this ML approach}

Typical MLIP model-construction schemes rely on training, validation, and test datasets. The training set is used to compute the model parameters; the validation set is used to refine the choice of the hyperparameters that define the model or the training conditions; finally, the test set is used to estimate the accuracy of the model's predictions. The three datasets contain information that is equivalent in a statistical sense, i.e., they are usually composed of structures chosen at random from a single batch of data. (A rule of thumb is to use at least 60~\% of the data for training, at least 10~\% for validation, and at least 10~\% for testing.) The approach employed here differs from this general protocol, which warrants some comments.

The only relevant hyperparameters in our construction scheme are the cut-offs (i.e., the choice of highest-order terms considered) in the expansion of the free energy as a function of polarization and strain components. Because of the mathematical simplicity of Landau potentials, it is trivial to work with cut-offs high enough as to make the universe of possible models essentially complete.

For a given choice of cut-offs (as, e.g., corresponding to the couplings listed in Table~\ref{tab:F}), we are left with the task of identifying the specific couplings that yield an optimum potential that is (1) as simple as possible, (2) as accurate as required, and (3) as reliable (predictive) as possible. This can be considered a second-level choice of hyperparameters. As explained above, we deal with this task automatically by monitoring the predictive error $\mathsf{e}$ (Eq.~(\ref{eq:e})) and, typically, selecting the simplest model that yields $\mathsf{e}$ below a certain threshold. This procedure thus constitutes a model validation, but with some key differences with respect to the usual approach. First, in our validation strategy, we do not employ a different dataset to evaluate the models by computing the (fitting) error $\mathsf{E}$. (As mentioned above, we do not find this strategy effective.) Rather, we consider the whole TS employed to fit the model parameters, but use the information in a different way. Specifically, we take advantage of the fact that all the states in the TS are stable equilibrium points, which implies that our model should be able to predict them as local free energy minima for the corresponding conditions of temperature, electric field and stress, as sketched in Fig.~\ref{fig:validation}. By defining the error function $\mathsf{e}$, which quantifies the accuracy of such a prediction, we implement a very demanding and informative validation criterion.

The error $\mathsf{e}$ corresponding to the validated model measures the maximum deviation we can expect when we use it in applications. Because the implemented validation procedure quantifies the predictive power of the models in an efficient manner that replicates the way in which the potentials will be used (that is, to compute equilibrium states in different conditions), we take it as a measure of the quality of our best model. Thus, within our scheme, the testing step is a byproduct of the validation step. 

Given the usefulness of the predictive error $\mathsf{e}$, one could wonder about the possibility of fitting the model's parameters by a direct minimization of this function. While possible in principle, this strategy suffers from a major disadvantage, namely, that we do not have a mathematical expression for $\mathsf{e}$ as a function of the model's parameters, let alone a formula for its derivatives with respect to said parameters. Hence, finding the minimum of $\mathsf{e}$ constitutes a challenging and computationally intensive numerical problem, to be contrasted with the simplicity of minimizing $\mathsf{E}$ (which boils down to the simple matrix operations in Eq.~(\ref{eq:solution})). Thus, minimizing $\mathsf{E}$ for fitting -- thus enabling the exploration of thousands of candidate models --, along with using $\mathsf{e}$ for validation (and testing), emerges as the most effective strategy.

In sum, we take advantage of the peculiarities of our TS and chosen models to implement validation and test steps that outperform the application of generic ML practices to our case. We contend that these ideas, or variations thereof, may be valuable and merit consideration in other contexts.

\subsection{Predictive ``third-principles'' simulations}

The present work is part of a longer-term effort to develop methods for mesoscale and macroscopic simulations of ferroic materials undergoing non-reconstructive phase transitions, a large family that includes many all-important ferroelectric and ferroelastic compounds. Traditionally, such simulations have relied on empirical Landau (macroscopic) and Ginzburg-Landau (mesoscopic) models. The present work addresses the question of whether such schemes may serve us as a basis to build potentials that can be improved systematically, capable of accounting for large sets of data that can be obtained from more basic and predictive simulations (as done here), from experiment, or eventually by combining both. The successful application to PTO, a less than trivial compound, suggests that polynomial Landau-type potentials are indeed a valid choice.

The next step in this program would be the generalization of the Landau models to situations where the order parameters can present a spatial variation (i.e., non-zero gradients), a key development that would enable the study of multidomain states. Such Ginzburg-Landau models would be fitted to data from atomistic simulations that capture thermodynamic properties of multidomain configurations. The key step in the procedure is to implement a mapping between the atomic configurations in the dataset and the order-parameter fields in the Ginzburg-Landau potential. This problem has already been discussed in the literature and field models for very complex compounds (e.g., PbZrO$_{3}$) have been derived in the limit of 0~K.\cite{schiaffino17,shapovalov23} While an application at finite temperatures involves some subtleties, the work plan is in principle clear. Additionally, it should be noted that for comparatively simple compounds -- like PbTiO$_{3}$ or BaTiO$_{3}$ -- it is conceivable to estimate the parameters of the most relevant gradient terms from experimental information, e.g., concerning the width of domain walls.\cite{tagantsev-book2010,meier-book2020} 

Having said this, we believe that the most important and least obvious step of this longer-term program pertains to the dynamical behavior of the order parameters, be they treated as macroscopic quantities or position-dependent fields. Indeed, the community has traditionally employed effective time-evolution equations based on Landau or Ginzburg-Landau potentials, of the form
\begin{equation}
    \mu \ddot{Q} + \gamma \dot{Q} = -\frac{\partial F}{\partial Q} \; ,
\end{equation}
where $\mu$ and $\gamma$ are effective inertial and damping constants, respectively, and $Q$ is a generic order parameter.\cite{chen07} Usually, such an equation is considered in the overdamped limit and is just used as a numerical tool to find minima of $F$. In such a context, the particular choice of $\mu$ and $\gamma$ is largely irrelevant, and authors assume $\mu=0$ and $\gamma=1$ in the vast majority of studies. However, the situation is quickly changing. Because of a growing interest in the response of complex electric and elastic textures to time dependent fields, including ultra-fast electric probes where the inertia of the polarization order parameter becomes relevant,\cite{yang20,chen24} the question of how to compute realistic values for the effective dynamic constants is starting to receive attention. Indeed, schemes to derive such parameters from more fundamental (atomistic) simulations are beginning to appear,\cite{liu21,wen25} and this problem will also be the focus of an upcoming article from our group. Let us stress that, as regards such effective dynamic constants, it may be possible to draw sensible values from experiment in some cases -- e.g., dielectric and infrared spectroscopy may offer relevant information about the $\gamma$ parameter for the polarization field.\cite{chen24} However, whenever we deal with a structural order parameter that cannot be easily probed experimentally -- e.g., the strain field, or the tilts of oxygen octahedra that control the structural behavior of most perovskite oxides\cite{chen18a} --, a theoretical approach to this problem becomes essential.

Thus, the present work is best understood as part of a larger effort to develop predictive simulation methods at mesoscopic and macroscopic scales. In earlier work, some of us introduced the term ``second principles'' to denote atomistic lattice-dynamical potentials formulated as polynomial expansions around physically relevant reference structures.\cite{wojdel13,escorihuelasayalero17,garciafernandez16} In the same spirit, we now propose the term ``third principles'' to describe the non-atomistic Landau and Ginzburg-Landau approaches discussed here. These third-principles methods also rely on a polynomial expansion around an appropriate reference state, which renders physically transparent and computationally light models. As shown in this work, such an approximation may be sufficiently accurate.

\section{Summary and conclusions}

In this work we introduce a scheme to build Landau-like free energy potentials suitable to describe non-reconstructive phase transitions, as those displayed by most ferroelectric and ferroelastic materials. The method is general and can be applied to any such compound -- from classic perovskite oxides like PbTiO$_{3}$, BaTiO$_{3}$ or BiFeO$_{3}$, to novel van der Waals ferroelectrics like hexagonal-BN\cite{wu21b} and many chalcogenides\cite{lipatov22} -- as long as one is able to generate relevant thermodynamic data (e.g., polarizations and strains as a function of temperature and applied fields) from a more fundamental -- atomistic -- theory. Further, the present approach is also suitable for use with experimental data.

This method complements statistical simulations based on machine-learned interatomic potentials, and it can be used -- in an essentially routine manner -- to derive Landau models from those. Our approach is useful to shed light on the behavior of complex materials, by deriving physically transparent Landau potentials that describe their phase transitions in terms of a few relevant couplings. Further, it enables a first necessary step towards predictive dynamical simulations, at the mesoscopic or macroscopic scales, based on Landau or Ginzburg-Landau models. Indeed, this work is part of a longer-term effort to develop what we call predictive ``third-principles'' methods.

We have demonstrated our approach with an application to ferroelectric perovskite PbTiO$_{3}$. By introducing an original model-validation step, which tests the predictive power of the potentials and yields a relevant numerical estimate of their accuracy, our machine-learning scheme makes it possible to automatically identify models that strike an optimum balance between accuracy, predictive power, and simplicity. We have shown that the identified optimal Landau potentials capture all the physical effects described in the training set (i.e., the non-reconstructive and weakly discontinuous structural transition of PbTiO$_{3}$; the system's response to applied electric fields and stresses) in a quantitative correct manner, while offering direct insights into the underlying physical couplings. Remarkably, our models also capture subtle physical phenomena -- concerning the non-trivial coupling between electric polarization and elastic strains in PbTiO$_{3}$ -- in a way that is qualitatively and quantitatively correct, while remaining physically transparent. Indeed, we show that the good performance of our optimum models extends to effects (e.g., the behavior of PbTiO$_{3}$ under peculiar elastic boundary conditions) that are not present in the training set used to fit them. In our view, this can be partly attributed to their overly simple and physically motivated form, and we contend it is probably an example of the predictive power of a broader class of physics-informed machine-learned potentials.

Work funded by the Luxembourg National Research Fund (FNR) through project C21/MS/15799044/FERRODYNAMICS. We thank A.~Gr{\"u}nebohm (Bochum) and Í.~Robredo (LIST) for fruitful discussions.

\end{document}